\documentclass[11pt,fleqn]{article}

\usepackage{amsmath,amsfonts,amssymb}
\usepackage{latexsym}
\usepackage{amsmath} 
\usepackage{amssymb} 
\usepackage{bm}
\usepackage{graphicx}
\usepackage{subfigure}

\begin{document} 


\title{Electromagnetic Mach principle}
\author{Eduardo Guendelman and Roee Steiner\\Physics Department, Ben-Gurion University, Beer-Sheva 84105, Israel}

\maketitle
\abstract{We will introduce a gauge model which an electromagnetic coupling constant and local mass are related to all the charge in the universe. we will use the standard Dirac action ,but where the mass and the electromagnetic coupling constant are a function of the sum of all the charge in the universe, which represent Mach principle for electromagnetic coupling constant.
The formalisation is not manifestly Lorentz invariant, however Lorentz invariance can be restored by performing a phase transformation of the Dirac field.}

\section{Introduction}
Mach's principle is well known, as a principle that relates a local problems to a non local problem. The original Mach principle \cite{Mach} is based on the claim that the inertial frames are influenced by the other celestial bodies.
In other words' every mass in the universe is influenced by all the other masses in the universe.
The Mach principle is still in a debate. 
In our article, we will try to show that there is a possibility to generalize Mach principle not only for mass but also for electromagnetic coupling constant in which the total charge brakes the locality of the problem (in the original Mach principle the mass broke the locality).

\section{The electromagnetic coupling constant as a function of all the charge in the universe}\label{per:charge}
We take the action of the Dirac equation (see for example ref \cite{basic})
\begin{equation}\label{eq:simple dirac action}
S=\int d^{4}x\, \bar{\psi}(\frac{i}{2}\gamma^{\mu}\stackrel{\leftrightarrow}{\partial}_{\mu}-eA_{\mu}\gamma^{\mu}-m)\psi
\end{equation}
  where $ \bar{\psi}=\psi^{\dagger}\gamma^0 $.However here we take the coupling constant to be proportional to the total charge (we will afterwards generalize and consider an arbitrary function of the total charge).

\begin{equation}\label{eq:def e}
e=\lambda\int\psi^{\dagger}(\vec{y},y^{0}=t_{0})\psi(\vec{y},y^{0}=t_{0})\, d^{3}y=\lambda\int \rho(\vec{y},y^{0}=t_{0})\, d^{3}y
\end{equation}

and we will show that physics does not depend on the time slice $ y^{0}=t_{0} $

so after the new definition of "e" the  action will be:

\begin{eqnarray}
S=\int d^{4}x\, \bar{\psi}(x)(\frac{i}{2}\gamma^{\mu}\stackrel{\leftrightarrow}{\partial}_{\mu}-m)\psi(x)\nonumber\\-\lambda(\int d^{3}y\, \bar{\psi}(\vec{y},y^{0}=t_{0})\gamma^{0}\psi(\vec{y},y^{0}=t_{0}))(\int d^{4}x\, \bar{\psi}(x)A_{\mu}\gamma^{\mu}\psi(x))
\end{eqnarray}

we can express the three dimensional integral as a four dimensional integral

\begin{equation}\label{eq:trick}
\int d^{3}y\, \bar{\psi}(\vec{y},y^{0}=t_{0})\gamma^{0}\psi(\vec{y},y^{0}=t_{0})= \int d^{4}y\, \bar{\psi}(y)\gamma^{0}\psi(y)\delta(y^{0}-t_{0}) 
\end{equation}

so finally the action will be

\begin{equation}
S=\int d^{4}x\, \bar{\psi}(x)(\frac{i}{2}\gamma^{\mu}\stackrel{\leftrightarrow}{\partial}_{\mu}-m)\psi(x)-\lambda(\int d^{4}x\, \bar{\psi}(x)A_{\mu}\gamma^{\mu}\psi(x))(\int d^{4}y\, \bar{\psi}(y)\gamma^{0}\psi(y)\delta(y^{0}-t_{0}))
\end{equation}

if we consider the fact that $ \frac{\delta\bar{\psi}_{a}(x)}{\delta\bar{\psi}_{b}(z)}=\delta^{4}(x-z)\,\delta_{ab} $ and $ \frac{\delta\psi(x)}{\delta\bar{\psi}(z)}=0 $ we set

\begin{eqnarray}
\frac {\delta S}{\delta\bar{\psi}(z)}=0=\int{\delta^{4}(x-z)(i\gamma^{\mu}\partial_{\mu}-m)\psi (x)}\,d^{4}x\,\nonumber\\-\lambda(\int d^{4}x\, \delta^{4}(x-z)A_{\mu}\gamma^{\mu}\psi(x))(\int d^{4}y\, \bar{\psi}(y)\gamma^{0}\psi(y)\delta(y^{0}-t_{0}))\nonumber\\-\lambda(\int d^{4}x\, \bar{\psi}(x)A_{\mu}\gamma^{\mu}\psi(x))(\int d^{4}y\, \delta^{4}(y-z)\gamma^{0}\psi(y)\delta(y^{0}-t_{0}))
\end{eqnarray}

so to accomplish our goal we need just to integrate the last equation, and then the expression will simplified to

\begin{eqnarray}
\frac {\delta S}{\delta\bar{\psi}(z)}=(i\gamma^{\mu}\partial_{\mu}-m)\psi(z)-\lambda(\int{\bar{\psi}(y)\gamma^{0}\psi(y)\delta(y^{0}-t_{0})}\,d^{4}y)A_{\mu}\gamma^{\mu}\psi(z)&\nonumber\\-\lambda(\int{\bar{\psi}(x)A_{\mu}\gamma^{\mu}\psi(x)}\,d^{4}x)\gamma^{0}\psi(z)\delta(z^{0}-t_{0})
\end{eqnarray}

which can be simplified more by the use of new definition  $ b_{e}=\lambda(\int{\bar{\psi}(x)A_{\mu}\gamma^{\mu}\psi(x)}\,d^{4}x) $ which is a constant, and by the definition in equation \ref{eq:def e} 

\begin{equation}\label{eq:motion e}
\frac {\delta S}{\delta\bar{\psi}(z)}=[i\gamma^{\mu}\partial_{\mu}-m-eA_{\mu}\gamma^{\mu}-b_{e}\gamma^{0}\delta(z^{0}-t_{0})]\psi(z)=0
\end{equation}

so we can see that the last term in the equation of motion \ref{eq:motion e} contains $ A^{GF}_{\mu}\gamma^{\mu} $ where $ A^{GF}_{\mu}=\partial_{\mu}\Lambda $ and $ \Lambda=b_{e}\theta(z^{0}-t_{0}) $ is a pure gauge field. so the solution of this equation is
\begin{equation}
\psi=e^{b_{e}\theta(z^{0}-t_{0})}\psi_{D} 
\end{equation}

  where $ \psi_{D} $ is the solution of the equation
  \begin{equation}
 [i\gamma^{\mu}\partial_{\mu}-m-eA_{\mu}\gamma^{\mu}]\psi_{D}=0 
  \end{equation}
  
    from which it follows that $ j^{\mu}=\bar{\psi}_{D}\gamma^{\mu}\psi_{D}=\bar{\psi}\gamma^{\mu}\psi $ and that $ Q=\int{d^{3}x\, j^{0}} $ is conserved, so it does not depend on the time slice .

\section{Mass as a function of all the charge in the universe}
we will show now that we can do the same thing as in paragraph \ref{per:charge} for the coupling constant of mass.
we take the action equation \ref{eq:simple dirac action} and take the mass to be equal to

\begin{equation}\label{eq:def m}
m=\lambda\int{\bar{\psi}\gamma^{0}\psi \,d^{3}y}
\end{equation}

we do the same thing like in paragraph \ref{per:charge} so we expend equation \ref{eq:def m} like we did in equation \ref{eq:trick} and divide by $ \delta\bar{\psi} $ and do the integration, so we will get the motion equation 

\begin{equation}\label{eq:motion m}
\frac {\delta S}{\delta\bar{\psi}(z)}=[i\gamma^{\mu}\partial_{\mu}-m-eA_{\mu}\gamma^{\mu}-b_{m}\gamma^{0}\delta(z^{0}-t_{0})]\psi(z)=0
\end{equation}

where $ b_{m}=\lambda\int{\bar{\psi}\psi \,d^{4}x} $. so again we have a gauge transformation, with the same conclusion like in paragraph \ref{per:charge} without violating Lorentz invariance.

\section{General coupling constant as a function of all the charge in the universe}

for a general coupling constant $"a"$ and a general function $ F(\bar{\psi},\psi) $ and a coupling constant "e" the action of the Dirac equation is

\begin{equation}\label{eq:general function action}
S=\int d^{4}x\, \bar{\psi}(\frac{i}{2}\gamma^{\mu}\stackrel{\leftrightarrow}{\partial}_{\mu}-eA_{\mu}\gamma^{\mu}-m)\psi+\int{d^{4}x[a F(\bar{\psi},\psi)]}
\end{equation}

we take the coupling constants "a" and "e" as an arbitrary function $g_{a}$ and $g_{e}$ of

 $Q=\int\psi^{\dagger}(\vec{y},y^{0}=t_{0})\psi(\vec{y},y^{0}=t_{0}) \,d^{3}y $ 

\begin{equation}\label{eq:general}
a=g_{a}(\int\psi^{\dagger}(\vec{y},y^{0}=t_{0})\psi(\vec{y},y^{0}=t_{0}) \,d^{3}y)=g_{a}(\int \rho(\vec{y},y^{0}=t_{0}) \,d^{3}y)
\end{equation}

\begin{equation}
e=g_{e}(\int\psi^{\dagger}(\vec{y},y^{0}=t_{0})\psi(\vec{y},y^{0}=t_{0}) \,d^{3}y)=g_{e}(\int \rho(\vec{y},y^{0}=t_{0}) \,d^{3}y)
\end{equation}

we put the last definition in equation \ref{eq:general function action} and expand equation \ref{eq:general} like we did in equation \ref{eq:trick}, we do a variation and use the fact that $ \frac{\delta\bar{\psi}_{a}(x)}{\delta\bar{\psi}_{b}(z)}=\delta^{4}(x-z)\,\delta_{ab} $ and $ \frac{\delta\psi(x)}{\delta\bar{\psi}(z)}=0 $ and do the integration as like we did in peragraph \ref{per:charge} so we get to the general motion equation
 
\begin{equation}\label{eq:motion general}
\frac {\delta S}{\delta\bar{\psi}(z)}=[i\gamma^{\mu}\partial_{\mu}-m-eA_{\mu}\gamma^{\mu}-b_{ae}\gamma^{0}\delta(z^{0}-t_{0})]\psi(z)+a(\frac{\partial F(\bar{\psi(z)},\psi(z))}{\partial\bar{\psi}(z)})=0
\end{equation}

where $ b_{ae}=\frac{\partial g_{a}(Q)}{\partial Q} \int{d^{4}x\,F(\bar{\psi},\psi)}+\frac{\partial g_{e}(Q)}{\partial Q}\int{d^{4}x\,\bar{\psi}\gamma_{\mu}\psi A^{\mu}} $
so we get that any coupling constant in this form can be a function of the charge in the universe without violating Lorentz invariance.

\section{Conclusion}
We have constructed a model that showed that we can make a new "Mach principle" for charge, in which all the electromagnetic coupling constant are a function of the total charge in the universe.

\begin{equation}
Q=\int\psi^{\dagger}(\vec{y},y^{0}=t_{0})\psi(\vec{y},y^{0}=t_{0}) \,d^{3}y=\int \rho(\vec{y},y^{0}=t_{0}) \,d^{3}y
\end{equation}

We have showed that Lorentz invariance can be restored by just a phase transformation of the Dirac field.
This shows that coupling constants can be taken not just as a given number but as a global function that depends on the global state of the universe in particular on the total charge of the universe, in the context of a consistent formalism.
This is an explicit realization of something similar to the Mach principle, but now applied to charge rather than to mass.

\end{document}